\documentclass[twocolumn,showpacs,preprintnumbers,amsmath,amssymb]{revtex4}

\usepackage{graphicx}
\usepackage{dcolumn}
\usepackage{bm}
\begin{document} 
\title{Antiferromagnetic Phases of the Kondo Lattice}
\author{R. Eder$^1$, K. Grube$^1$, P. Wr\'obel$^2$}
\affiliation{$^1$ Karlsruhe Institute of Technology,
Institut f\"ur Festk\"orperphysik, 76021 Karlsruhe, Germany\\
$^2$Institute for Low Temperature and Structure Research,
             P.O. Box 1410, PL-50950 Wroc{\l}aw 2, Poland}
\date{\today}
\begin{abstract}
We discuss the paramagnetic and N\'eel-ordered phases of the Kondo lattice 
Hamiltonian on the 2D square lattice by means of bond Fermions. 
In the doped case we find two antiferromagnetic solutions, the first one 
with small ordered moment, heavy bands, and an antiferromagnetically folded 
large Fermi surface - i.e. including the localized spins - the second one with 
large ordered moment, light bands and an antiferromagnetically folded 
conduction electron-only Fermi surface. The zero temperature phase diagram as 
a function of Kondo coupling and conduction electron density shows
first and second order transition lines between the three different phases
and agrees qualitatively with previous numerical studies. We compare to
experiments on CeRh$_{1-x}$Co$_x$In$_5$ and find qualitative agreement.
\end{abstract} 
\maketitle
\section{Introduction}
Heavy Fermion compounds are a much studied class of materials in the field of
strongly correlated electron systems. Among the many
phenomena observed in these compounds is a variety of phase transitions
between different magnetically ordered and nonmagnetic phases which occur as 
a function of temperature, pressure, alloying, or magnetic field.
Often the transition temperature can be driven to zero Kelvin
by varying some experimental parameter
resulting in quantum critical points and superconductivity\cite{Review}.
The simplest model believed to be able to describe these compounds
is the Kondo lattice model, which is obtained from the more
realistic periodic Anderson model by means of the Schrieffer-Wolff 
transformation\cite{SchriefferWolff} and describes a single conduction band
coupled to a periodic array of loaclized spins by spin-exchange,
\begin{eqnarray}
H&=&\sum_{{\bf k},\sigma}\;\epsilon_{\bf k}\;c_{{\bf k},\sigma}^\dagger c_{{\bf k},\sigma}^{}
+ J\sum_i \;\vec{S}_i\cdot\vec{\sigma}_i.
\label{kondola}
\end{eqnarray}
Thereby each unit cell $i$ is assumed to
contain one conduction band (or $c$) orbital and one
localized (or $f$) orbital, the operators
$c_{i,\sigma}^\dagger$ and $f_{i,\sigma}^\dagger$
create an electron with z-spin $\sigma$ in these.
Moreover, $\vec{\sigma}_i= \frac{1}{2}\;c_{i\sigma}^\dagger \;
\vec{\tau}_{\sigma \sigma'}\; c_{i\sigma'}^{}$ where
$\vec{\tau}$ is the vector of Pauli matrices whereas $\vec{S}_i$ denotes the
spin operator of the localized electrons and $\epsilon_{\bf k}$ is the
dispersion relation of the conduction band.\\
It is widely believed that the magnetic phase transitions are the consequence
of a competition between the Kondo-effect and the 
Ruderman-Kittel-Kasuya-Yoshida (RKKY) interaction between the
localized electrons\cite{Doniach}.
The essence of the Kondo effect is the formation of singlets between a
localized electron and a conduction electron, leading to a vanishing of
the expectation value $\langle \vec{S}_i\rangle$, whereas the
RKKY-interaction - or any other mechanism favouring magnetic order such
as a magnetic field -
favours a nonvanishing $\langle \vec{S}_i \rangle$.
Accordingly, there is an inherent frustration in the Kondo lattice
model and slight perturbations may tilt the balance and
induce a phase transition. Moreover, for a lattice of localized spins 
the Kondo effect leads to a Fermi surface volume to which the localized
electrons contribute as if they were itinerant. The transitions between 
`Kondo-dominated' and `RKKY-dominated' phases therefore are often accompanied 
by a reconstruction of the Fermi surface which goes beyond the simple 
generation of umklapps but rather changes the volume of the Fermi surface 
by an amount corresponding to half of an electron per spin and $f$-site.\\
While the single impurity Kondo problem can be solved exactly the
Kondo lattice Hamiltonian is less well understood.
Following the work of Yoshimori and Sakurai\cite{YoshimoriSakurai},
Lacroix and Cyrot\cite{LacroixCyrot,Lacroix} 
studied the Kondo lattice in mean field theory.
For a band with a constant density of states in the range $[-D:D]$ and 
electron densities close to $n_c=1$ Lacroix and Cyrot found a paramagnetic
and a N\'eel ordered solution. At $T=0$ 
the antiferromagnetic phase thereby
turns out to have lower energy than the paramagnetic one for $J\le D/2$.
Its ordered $f$-moment always has the maximum value of
$1/2$\cite{LacroixCyrot} so that the phase transition is 1$^{st}$ order.
Various other mean-field studies of the Kondo lattice
were performed since then\cite{ZhangYu,Senthil,ZhangSuLu,Nilsson,Asadzadeh} 
but it appears to be difficult to reproduce the phase diagrams of Heavy Fermion 
compounds by this approach.\\
Since then the model was also studied by renormalization 
group\cite{renormalization} and extended dynamical mean 
field theory (see Ref. \cite{EDMFT} for a recent review) 
and a global phase diagram was outlined\cite{EDMFT,Global}.
Additional work has focussed on the properties of quasiparticles near the 
quantum critical points\cite{woelfle}
and the phenomenological two-fluid model was 
proposed (for a recent review see Ref. \cite{twofluid}).\\
In the present manuscript we present an approximation in terms of bond
Fermions which reproduces a few results obtained previously only by 
numerical methods, such as the phase diagram containing two antiferromagnetic 
and a paramagnetic phase. As will be discussed below this phase diagram also 
qualitatively reproduces some experimental results on Heavy Fermion compounds.
\section{Method of Calculation}
\begin{figure}
\includegraphics[width=0.7\columnwidth]{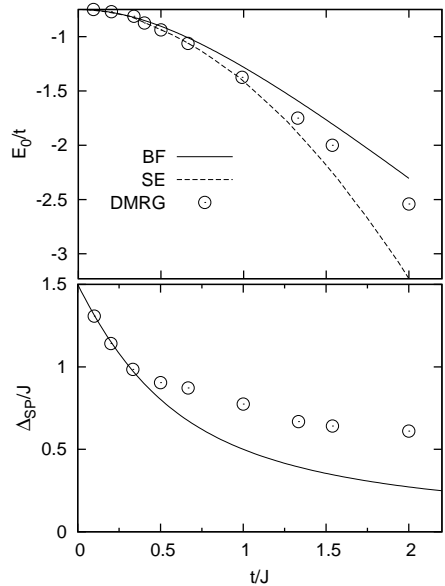}
\caption{\label{fig1} 
Ground state energy $E_0/t$ per site (top) and single particle gap 
$\Delta_{QP}/J$ versus $t/J$ for the  1D Kondo insulator.
The line is the result from the bond Fermion calculation the symbols
are DMRG results by Yu and White\cite{YuWhite}. The curve labeled SE
in the top panel is the ground state energy obtained by series
expansion\cite{series}.}
\end{figure}

In the following we consider the Hamiltonian (\ref{kondola})
on a 2D square lattice of $N$ unit cells,
the number of conduction electrons is $N_c$ and 
$n_c=N_c/N=1-\delta$.
For the dispersion of the conduction band, $\epsilon_{\bf k}$,
we assume a tight-binding form with hopping integrals $-t$ between
nearest and $t_1$ between second nearest neigbors i.e.
\[
\epsilon_{\bf k}=-2t\left(\;\cos(k_x)+\cos(k_y)\;\right) +4t_1\;
\cos(k_x)\;\cos(k_y).
\]
The calculation to be outlined below may be
viewed as a Fermionic version of the bond operator theory
proposed by Sachdev and Bhatt to describe Bosonic spin fluctuations
in quantum spin systems\cite{SachdevBhatt} which was applied
successfully to spin ladders \cite{Gopalan}.
In deriving the Fermionic version we follow Ref. \cite{Oana}, see also 
Ref. \cite{JureckaBrenig} for a more rigorous derivation.
In the limit $t, t_1\rightarrow 0$ and $N_c=N$ the
ground state is a product of singlets
\begin{eqnarray}
|\Psi_0\rangle &=&\prod_{j=1}^N\;s_j^\dagger\; |0\rangle, \nonumber \\
s_j^\dagger&=&\frac{1}{\sqrt{2}}\;
(c_{j\uparrow}^\dagger f_{j\downarrow}^\dagger - c_{j\downarrow}^\dagger f_{j\uparrow}^\dagger).
\label{vacuum_0}
\end{eqnarray}
The energy of this state is $-N e_0$ with $e_0=\frac{3}{4}J$. Switching on 
nonvanishing hopping integrals produces {\em charge fluctuations}, e.g.
an electron with spin $\sigma$ can be transferred from some cell $m$
to another cell $n$, resulting in a state
with three electrons in $n$ and a single electron in $m$.
In subsequent steps either the surplus electron or the hole may propagate to 
other sites or additional electron-hole pairs may be generated.
In Ref. \cite{Oana} the product of singlets (\ref{vacuum_0}) was considered 
as the vacuum $|vac\rangle$ for charge fluctuations which themselves
were described as effective Fermions (here we will call them `bond Fermions')
There are hole-like and electron-like bond Fermions, and states containing a 
single of these correspond to states of the true Kondo lattice as follows:
\begin{eqnarray}
b_{i\sigma}^\dagger |vac\rangle &\rightarrow& 
c_{i\uparrow}^\dagger c_{i\downarrow}^\dagger\; f_{i\sigma}^\dagger \;
\prod_{j\ne i}\;s_j^\dagger\;|0\rangle,\nonumber \\
a_{i\sigma}^\dagger |vac\rangle &\rightarrow& 
f_{i\sigma}^\dagger \;
\prod_{j\ne i}\;s_j^\dagger\;|0\rangle.
\label{singles}
\end{eqnarray}
The generalization to states with more than one bond Fermion
is self-evident, the only requirement being that the factors of
$c_{i\uparrow}^\dagger c_{i\downarrow}^\dagger\; f_{i\sigma}^\dagger$
and $f_{j\sigma}^\dagger$ in the Kondo-lattice states
be in the same order
as the $b_{i\sigma}^\dagger$ and $a_{j\sigma}^\dagger$ in the bond Fermion states.
Since a unit cell with either one or three electrons has an exchange
energy of $0$ we ascribe an energy of formation of $+e_0$ to each bond Fermion.
Operators for the bond Fermions are obtained by demanding that their
matrix elements between bond Fermions states be identical to those of
the physical operator between the respective translated states.
Due to the product nature of states like (\ref{singles}) this
is usually easy to achieve.
For example, the electron annihilation operator becomes
\begin{eqnarray}
c_{i,\sigma}&=&\frac{1}{\sqrt{2}}\left(\;sign(\sigma)\; a_{i,\bar{\sigma}}^\dagger 
- b_{i,\sigma}^{}\;\right).
\label{annihilation}
\end{eqnarray}
Fourier transformation gives the representation
of $c_{{\bf k},\sigma}$ and inserting this into the kinetic energy
$\sum_{{\bf k},\sigma}\;\epsilon_{\bf k}\;c_{{\bf k},\sigma}^\dagger c_{{\bf k},\sigma}^{}$
the Hamiltonian becomes\cite{Oana}:
\begin{eqnarray} 
H &=& \sum_{{\bf k},\sigma}
[\;(\frac{\epsilon_{\bf k}}{2}+ e_0)\;
b_{{\bf k},\sigma}^\dagger b_{{\bf k},\sigma}^{} +
(\frac{\epsilon_{\bf k}}{2}-e_0)\; 
a_{-{\bf k},\bar{\sigma}}^{} a_{-{\bf k},\bar{\sigma}}^\dagger 
\nonumber \\
&& -sign(\sigma)\; \frac{\epsilon_{{\bf k}}}{2}\;
(\;b_{{\bf k},\sigma}^\dagger a_{-{\bf k},\bar{\sigma}}^\dagger  + H.c.)\;] + N\; e_0.
\label{stcham}
\end{eqnarray}
The positive sign of the additive constant is not a misprint -
rather, this is the sum of the energy of the vacuum state (\ref{vacuum_0}), 
$-Ne_0$, and a term $+2Ne_0$ obtained by inverting $2N$ products of Fermion
operators
$e_0\;a_{-{\bf k},\bar{\sigma}}^\dagger a_{-{\bf k},\bar{\sigma}}^{}=e_0 -
e_0 \;a_{-{\bf k},\bar{\sigma}}^{} a_{-{\bf k},\bar{\sigma}}^\dagger$.
In using the Hamiltonian (\ref{stcham}) we are making two approximations:
first, the possibility that a unit cell containing two electrons
is in a triplet state is neglected. This means we neglect Bosonic spin
excitations and their coupling to the Fermionic charge fluctuations. Second,
states like $a_{i,\sigma}^\dagger b_{i,\sigma'}^\dagger |vac\rangle$
or $a_{i,\uparrow}^\dagger a_{i,\downarrow}^\dagger |vac\rangle$
where two bond Fermions occupy the same unit cell obviously
are meaningless so that the bond Fermions have to obey a hard core
constraint - which we neglect in the following. This issue is
discussed in detail in section V.\\
Since the vacuum state (\ref{vacuum_0}) contains $2N$ electrons 
{\em including the localized electrons} and since adding
an $a^\dagger$-Fermion (a $b^\dagger$-Fermion) decreases
(increases) the electron number by one, the total
number of electrons is
\begin{eqnarray}
N_e &=& 2N + \sum_{{\bf k},\sigma}\;b_{{\bf k}\sigma}^\dagger b_{{\bf k}\sigma}^{}
- \sum_{{\bf k},\sigma}\;a_{{\bf k}\sigma}^\dagger a_{{\bf k}\sigma}^{}
\nonumber \\
&=& \sum_{{\bf k},\sigma}\;\left( b_{{\bf k}\sigma}^\dagger b_{{\bf k}\sigma}^{}
+a_{-{\bf k}\bar{\sigma}}^{}  a_{-{\bf k}\bar{\sigma}}^\dagger \right).
\label{totel}
\end{eqnarray}
An extra complication - discussed in detail in Ref. \cite{Oana} -
is the following: after tuning the electron number to any prescribed value
$N'$ by adding the term $-\mu (N_e-N') $ to the
Hamiltonian and adjusting $\mu$, the ${\bf k}$-integrated conduction electron
momentum distribution function
$n_{\bf k}=\langle c_{{\bf k}\sigma}^\dagger c_{{\bf k}\sigma}^{}\rangle$ -
which equals the ${\bf k}$- and $\omega$ integrated photoemission weight -
in general is not equal to $N_c=N'-N$. As proposed
in Ref. \cite{Oana} we resolve this problem by enforcing the
equalitiy of $N_c$ and ${\bf k}$-integrated $n_{\bf k}$
via an additional Lagrange multiplier $\lambda$, i.e. we add the term
$-\lambda( \sum_{{\bf k},\sigma}c_{{\bf k}\sigma}^\dagger c_{{\bf k}\sigma}^{} - N_c)$
to $H$ and adjust $\lambda$.
This amounts to replacing $\epsilon_{{\bf k}}\rightarrow \epsilon_{{\bf k}}
- \lambda$ in (\ref{stcham}).
The Hamiltonian (\ref{stcham}) then can be solved by a
unitary transformation
\begin{eqnarray*}
\gamma_{{\bf k},1,\sigma}^\dagger&=&\;\;u_{\bf k} \;b_{{\bf k},\sigma}^\dagger
+ sign(\sigma)\;v_{\bf k} \;a_{-{\bf k},\bar{\sigma}}^{},\nonumber \\
\gamma_{{\bf k},2,\sigma}^\dagger &=&- sign(\sigma)\;v_{\bf k} \;b_{{\bf k},\sigma}^\dagger
+ u_{\bf k} \;a_{-{\bf k},\bar{\sigma}}^{}.
\end{eqnarray*}
In terms of the quasiparticle operators $\gamma_{{\bf k},\nu,\sigma}^\dagger$
the electron number (\ref{totel}) becomes
\begin{eqnarray*}
N_e &=&\sum_{{\bf k},\sigma}\;\sum_{\nu=1}^2\;
\gamma_{{\bf k},\nu,\sigma}^\dagger \gamma_{{\bf k},\nu,\sigma}^{}.
\end{eqnarray*}
While all bond Fermion basis states such as (\ref{singles})
have exactly one $f$-electron per unit cell so that the $f$-electrons are 
perfectly localized, the Fermi surface volume therefore is such as if the 
$f$-electrons were itinerant. 
The quasipartile dispersion and conduction electron
momentum distribution become (for $n_c\le 1$):
\begin{eqnarray*}
E_{{\bf k},\pm}&=& \frac{1}{2}\left( (\epsilon_{\bf k}-\lambda) \pm 
\sqrt{\left(\epsilon_{\bf k}-\lambda\right)^2 + 4e_0^2} \right) -\mu.
\nonumber \\
n_{\bf k}&=&\frac{1}{2}\left(1 -
\frac{\epsilon_{\bf k}-\lambda}{\sqrt{\left(\epsilon_{\bf k}-\lambda\right)^2 + 4e_0^2}}\right)\;\Theta(\mu-E_{{\bf k},-}).
\end{eqnarray*}
For $J/t\rightarrow 0$ the band structure approaches
the noninteracting $\epsilon_{\bf k}$ plus a
dispersionless band at energy $\lambda$ whereas
$n_{\bf k}\propto \Theta(\lambda-\epsilon_{\bf k})$.
In this limit both $\lambda$ and $\mu$ must be set
equal to the noninteracting
chemical potential whence the total energy becomes that
of the free Fermi sea.
For $J/t\rightarrow \infty$ we have two bands at $\pm e_0+ O(t)$.
For $N_c< N$ the chemical potential cuts into the lower of these 
two bands, resulting in a total energy of $-N_c e_0 + O(t)$.
Again, this is the correct limiting behaviour because for 
$J/t\rightarrow \infty$ the ground state
has $N_c$ singlets and $N-N_c$ mobile $c$-vacancies
which contribute an energy $\propto \delta\cdot t$.
The ground state energy obtained from the bond Fermion calculation
accordingly interpolates between these two exactly known limiting
values. Figure \ref{fig1} compares some results 
obtained in this way for a 1D chain with $n_c=1$ - i.e. the Kondo
insulator - and $t_1=0$ to the Density Matrix Renormalization Group 
(DMRG) results of Yu and White\cite{YuWhite}.
Since particle-hole symmetry requires $\lambda=\mu=0$ in this case,
the bond Fermion calculation gives
the ground state energy per site $E_0$ and single particle gap
$\Delta_{SP}=E_{k=0,+}-E_{k=\pi,-}$ as
\begin{eqnarray*}
E_0 &=& e_0- \frac{1}{\pi}\;
\int_{-\pi}^{\pi}\;dk\;\sqrt{t^2\cos^2(k)+e_0^2},\nonumber \\
\Delta_{SP}&=&2(\sqrt{e_0^2 + t^2}-t).
\end{eqnarray*}
This agrees reasonably well with DMRG for $t/J<1$ whereas the 
agreement is less satisfactory for $t/J>1$. 
The panel for the energy also shows the series expansion results
from Ref. \cite{series}.
The rather simple analytic bond Fermion estimate is as good
as the series expansion result for $t/J<1$ but appears to be
closer to the numerical result for $t/J>1$.\\
\begin{figure}
\includegraphics[width=0.6\columnwidth]{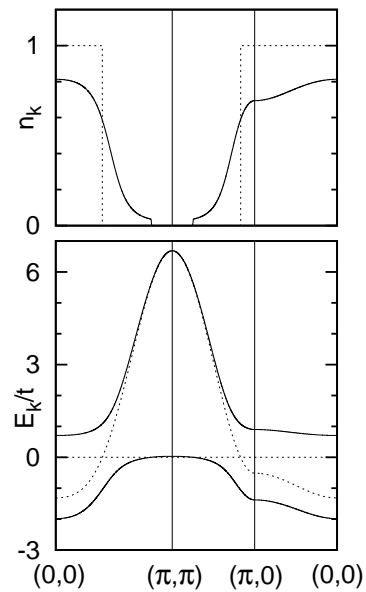}
\caption{\label{fig2} 
Conduction electron momentum distribution (top) and band structure (bottom)
for the paramagnetic phase of the 2D Kondo lattice Hamiltonian
with $n_c=0.9$, $J/t=1.4$ and $t_1/t=0.4$. The dashed lines show the
respective quantity for noninteracting electrons.}
\end{figure}
The band structure in two dimensions for finite $J/t$ - shown in in 
Figure \ref{fig2} -
has the familiar `hybridization gap' form, $n_{\bf k}$ has a sharp but 
continuous 
drop at the noninteracting Fermi surface and a tiny discontiuity
at the actual Fermi momentum. It should be stressed that 
since no self-consistency equation is solved in the bond Fermion
calclation the energy scale of the single impurity
Kondo temperature, $k_B T_K = D\; exp(1/{\rho_0 J})$
(with $\rho_0$ the density of states at the Fermi level
and $D$ the conuction electron bndwidth)
does not appear anywhere in the present clculation - 
if this plays a role also in the lattice case the bond Fermion
calculation is too crude to reproduce it.\\
We now generalize the calculation to a magnetically ordered phase.
In (\ref{vacuum_0}) and (\ref{singles})
we replace $s_j^\dagger\rightarrow \tilde{s}_j^\dagger$ where
\begin{eqnarray}
\tilde{s}_j^\dagger &=&\cos(\Theta)\; s_j^\dagger 
+ e^{i{\bf Q}\cdot {\bf R}_j}\sin(\Theta)\; t_{j,z}^\dagger\nonumber \\
 t_{j,z}^\dagger&=&\frac{1}{\sqrt{2}}\;
(c_{j\uparrow}^\dagger f_{j\downarrow}^\dagger + c_{j\downarrow}^\dagger f_{j\uparrow}^\dagger),
\label{vacuum_1}
\end{eqnarray}
with ${\bf Q}=(\pi,\pi)$ the antiferromagnetic wave vector
and the angle $\Theta$ will be determined subsequently my minimizing the energy.
The new vacuum state may be viewed as a condensate of Bosonic 
$z$-like triplets\cite{SachdevBhatt,JureckaBrenig} with momentum 
${\bf Q}$ in the  pure `singlet background' considered above
and has energy $-N\tilde{e}_0$ with
$\tilde{e}_0 = \frac{3J}{4}\cos^2(\Theta) - \frac{J}{4}\sin^2(\Theta)$.
For $\Theta=0$ we recover the original paramagnetic vacuum state
whereas for $\Theta=\frac{\pi}{4}$ we have the fully polarized
N\'eel state with opposite ordered moment for $c$ and $f$ electrons.
The energy of a bond Fermion now is $\tilde{e}_0$ and
instead of (\ref{annihilation}) we find
\begin{eqnarray*}
c_{i,\uparrow}&=& \;\;\;a_+ a_{i,\downarrow}^\dagger - a_- b_{i,\uparrow},\nonumber \\
c_{i,\downarrow}&=& -a_- a_{i,\uparrow}^\dagger - a_+ b_{i,\downarrow},\nonumber \\
&&\\
a_\pm &=&\frac{\cos(\Theta)\pm e^{i{\bf Q}\cdot {\bf R}_i}\sin(\Theta)}{\sqrt{2}}.
\end{eqnarray*}
We introduce the sublattices $A$ and $B$ whereby $A$ contains $(0,0)$
and accordingly introduce two species of bond Fermions, 
e.g. $b_{i,\sigma}^\dagger$ for $i\in A$ and
$\tilde{b}_{j,\sigma}^\dagger$ for $j\in B$ (and analogously for 
the $a^\dagger$'s). The Fourier transforms of the electron operators are
\begin{eqnarray*}
c_{{\bf k},\uparrow}&=&-c_- b_{{\bf k},\uparrow}
- c_+ \tilde{b}_{{\bf k},\uparrow}
+ c_- \tilde{a}_{-{\bf k},\downarrow}^\dagger,
+c_+ a_{-{\bf k},\downarrow}^\dagger,\nonumber \\
c_{{\bf k}+{\bf Q},\uparrow}&=&-c_- b_{{\bf k},\uparrow}
+ c_+ \tilde{b}_{{\bf k},\uparrow}
- c_- \tilde{a}_{-{\bf k},\downarrow}^\dagger,
+c_+ a_{-{\bf k},\downarrow}^\dagger,\nonumber \\
c_{{\bf k},\downarrow}&=&-c_+ b_{{\bf k},\downarrow}
- c_- \tilde{b}_{{\bf k},\downarrow}
 -c_+ \tilde{a}_{-{\bf k},\uparrow}^\dagger,
-c_- a_{-{\bf k},\uparrow}^\dagger,\nonumber \\
c_{{\bf k}+{\bf Q},\downarrow}&=&-c_+ b_{{\bf k},\downarrow}
+ c_- \tilde{b}_{{\bf k},\downarrow}
+c_+ \tilde{a}_{-{\bf k},\uparrow}^\dagger,
-c_- a_{-{\bf k},\uparrow}^\dagger,\nonumber \\
c_\pm&=&\frac{1}{2}(\cos(\Theta)\pm \sin(\Theta)),
\end{eqnarray*}
where ${\bf k}$ denotes a wave vector 
in the antiferromagnetic Brillouin zone
(AFBZ). By again inserting the above representations of
$c_{{\bf k},\sigma}$ into the kinetic energy we obtain the Hamiltonian.
Introducing the column vector
${\bf v}_\sigma({\bf k})=(b_{{\bf k},\sigma},\tilde{b}_{{\bf k},\sigma},
\tilde{a}_{-{\bf k},\bar{\sigma}}^\dagger,a_{-{\bf k},\bar{\sigma}}^\dagger)^T$
$H$ becomes
\begin{eqnarray*}
H&=&\sum_{{\bf k}\in AFBZ}\;\sum_\sigma\;
{\bf v}_\sigma^\dagger({\bf  k})\;H_\sigma({\bf  k})\;{\bf v}_\sigma({\bf  k})
+ N\;\tilde{e}_0,\nonumber \\
&&\nonumber \\
H_\sigma({\bf  k})&=&H_J + (\epsilon^{(+)}_{\bf k}-\lambda)\;W^{(+)}_\sigma +
\epsilon^{(-)}_{\bf k}\;W^{(-)}_\sigma,\nonumber \\
&&\nonumber \\
H_J &=&diag(\tilde{e}_0,\tilde{e}_0,-\tilde{e}_0,-\tilde{e}_0),\\
&&\nonumber \\
\epsilon^{(\pm)}_{\bf k}&=&\frac{1}{2}\left(\;\epsilon_{\bf k} \pm 
\epsilon_{{\bf k}+{\bf Q}}\;\right),
\end{eqnarray*}
and the matrices $W$ are given by
\begin{eqnarray*}
W_\sigma^{(+)} &=&\left(\begin{array}{c c c c}
\alpha_\mp,&0,&0,&\mp \beta\\
0,&\alpha_\pm,&\mp \beta,&0\\
0,&\mp \beta,&\alpha_\mp,&0\\
\mp \beta,&0,&0,&\alpha_\pm
\end{array}\right),\nonumber\\
&&\nonumber \\
W_\sigma^{(-)} &=&\left(\begin{array}{c c c c}
0,& \beta, & \mp \alpha_\mp, & 0\\
\beta, & 0,& 0,& \mp \alpha_\pm\\
\mp \alpha_\mp, &  0,& 0,& \beta\\
0,& \mp \alpha_\pm,& \beta,& 0
\end{array}\right).
\end{eqnarray*}
Here the upper (lower) sign on the respective right hand side
refers to $\sigma=\uparrow$ ($\sigma=\downarrow$) and
$\alpha_\pm=\frac{1}{2}\;\left(1\pm\sin(2\Theta)\right)$
and $\beta = \frac{1}{2}\;\cos(2\Theta)$.
The number of electrons is
\begin{eqnarray*}
N_e &=&\sum_{{\bf k}\in AFBZ}\;\sum_\sigma\;\sum_{\nu=1}^4\;
\gamma_{{\bf k},\nu,\sigma}^\dagger \gamma_{{\bf k},\nu,\sigma}^{}.
\end{eqnarray*}
Again, the electron number is obtained by filling the $4$
bands as if the $f$-electrons were intinerant and did participate
in the Fermi surface volume and the value of $\langle N_e\rangle $ is 
fixed by tuning the chemical potential $\mu$.
The parameter $\lambda$ is again adjusted to match real-space
count and ${\bf k}$-space count for the conduction electrons i.e.
\begin{eqnarray*}
\sum_{{\bf k}\in AFBZ}\;\langle\;n_{\bf k} + n_{{\bf k}+{\bf Q}}\;\rangle&=& N_e-N,
\end{eqnarray*}
whereby
\begin{eqnarray*}
n_{\bf k} + n_{{\bf k}+\bf{Q}}&=& \sum_\sigma \;{\bf v}_\sigma^\dagger \;W^{(+)}_\sigma\;
{\bf v}_\sigma^{}.
\end{eqnarray*}
In this way, the energy $\langle H \rangle$
can be calculated as a function of the angle $\Theta$
and minimized with respect to $\Theta$.
\section{Results}
\begin{figure}
\includegraphics[width=0.7\columnwidth]{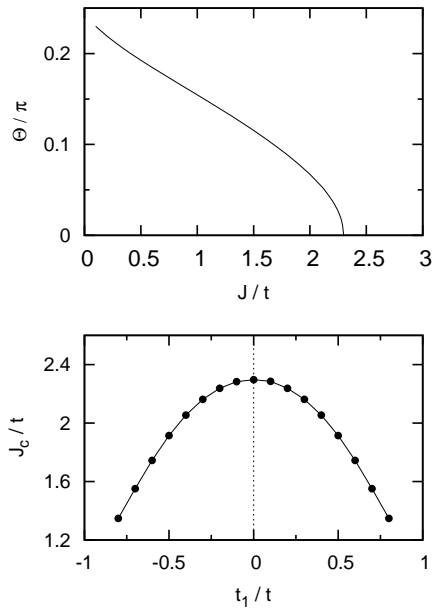}
\caption{\label{fig3} 
Top: Optimal angle $\Theta$ at $n_c=1$ and $t_1/t=0$ versus $J/t$. Bottom:
Values of $J_c/t$ where the phase transition occurs as a function
of $t_1/t$.}
\end{figure}
\begin{figure}
\includegraphics[width=0.7\columnwidth]{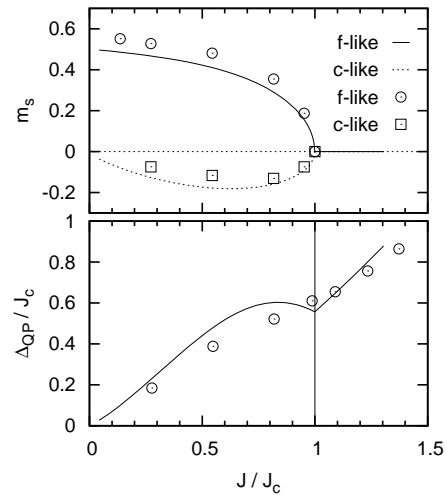}
\caption{\label{fig4} 
Top: Ordered moment from the bond Fermion calculation (lines) versus 
$J/J_c$ compared to QMC (symbols)\cite{Assaad}.\\
Bottom: Single particle gap $\Delta_{QP}$
from the bond Fermion calculation (line) versus
 $J/J_c$  compared to QMC (symbols) \cite{Assaad}.}
\end{figure}
\begin{figure}
\includegraphics[width=0.9\columnwidth]{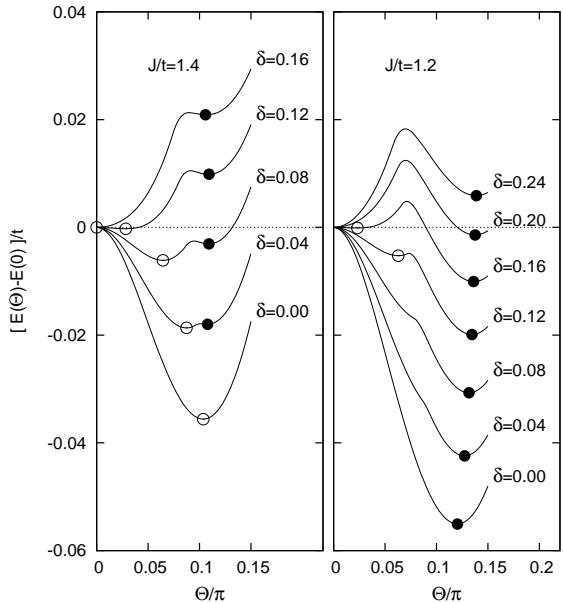}
\caption{\label{fig5} 
Ground state energy versus $\Theta$ for different 
$\delta$ and $t_1/t=0.4$.}
\end{figure}
\begin{figure}
\includegraphics[width=0.7\columnwidth]{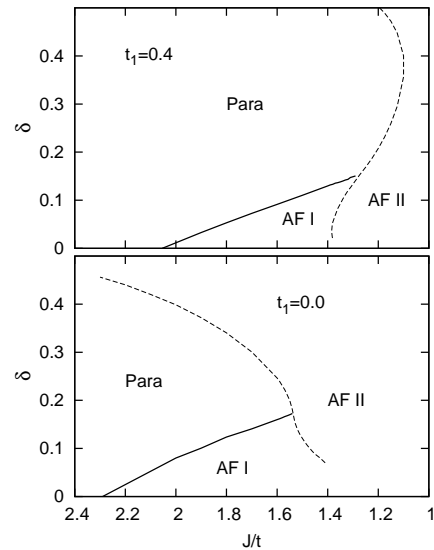}
\caption{\label{fig6} 
Phase diagram of the Kondo lattice on a 2D square lattice
as a function of Kondo coupling $J$ and hole density $\delta$. }
\end{figure}
\begin{figure}
\includegraphics[width=0.7\columnwidth]{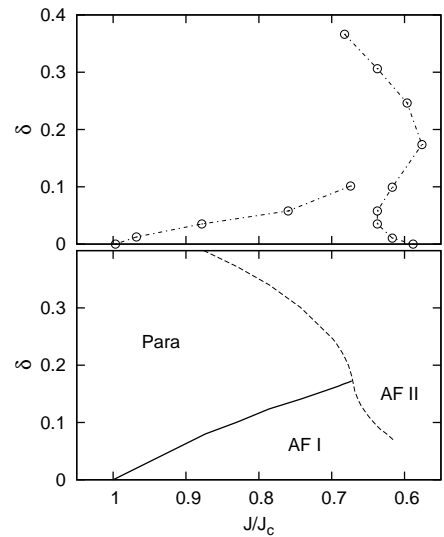}
\caption{\label{fig7} 
Phase diagram for $t_1/t=0$
obtained by VMC\cite{WatanabeOgata} (top)
compared to the phase diagram from bond Fermion theory (bottom).}
\end{figure}
\begin{figure}
\includegraphics[width=0.9\columnwidth]{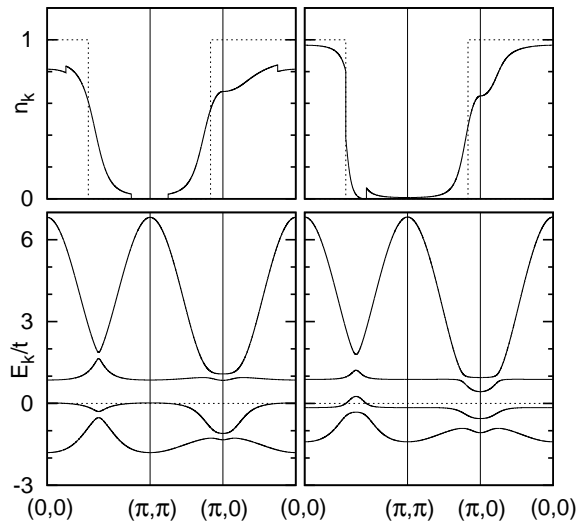}
\caption{\label{fig8} 
Top: Momentum distribution (top) and band structure
(bottom) for the AF-I phase (left) and AF-II phase(right).
In both panels $t_1/t=0.4$, $\delta=0.1$ whereas $J/t=1.4$ for
the  AF-I phase and $J/t=1.2$ for AF-II.}
\end{figure}
\begin{figure}
\includegraphics[width=0.7\columnwidth]{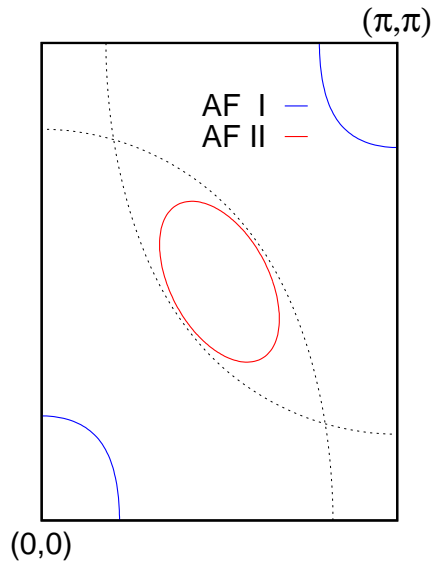}
\caption{\label{fig9} 
Fermi surface for the two antiferromagnetic phases
in Figure  \ref{fig8}. The dashed lines are the noninteracting
Fermi surface and its antiferromagnetic umklapp.}
\end{figure}

Figure \ref{fig3} shows the angle $\Theta$ which minimizes the
energy as a function of the ratio $J/t$ for
$n_c=1$ (i.e. the so-called Kondo insulator) and $t_1=0$.
At a certain $J_c\approx 2.29\; t$ $\Theta$ starts to deviate from
the value $\Theta=0$ which gives optimum energy at large $J/t$,
signalling a continuous - i.e. 2$^{nd}$ order - phase transition
to a magnetically ordered state (in the following $J_c$ will
always denote the value of $J$ below which antiferromagnetism sets in
for the Kondo insulator). For
$J/t\rightarrow 0$ $\Theta\rightarrow \frac{\pi}{4}$,
which means that the ordered $f$-moment
approaches its maximum. Figure \ref{fig3} also shows the value of
$J_c/t$ obtained for different values of $t_1/t$.
Switching on $t_1$ reduces 
$J_c/t$ whereby to good accuracy $J_c/t=2.29 - 1.50\;(t_1/t)^2$. This 
illustrates the change of the RKKY interaction due to the deformation of
the Fermi surface.\\
The exact value of $J_c/t=1.45$ for the case $t_1=0$ has been obtained by 
Assaad by Quantum Monte Carlo (QMC) calculation\cite{Assaad}.
The bond Fermion value is larger by a factor of $1.6$ which is on one hand
somewhat disappointing but on the other hand the energy difference
between the paramagnetic and antiferromagnetic phase is quite
small - see below - so that some deviation is to be expected.
Figure \ref{fig4} compares the ordered moment 
\[
m_\alpha = \frac{1}{2N}\;\sum_i \;e^{i{\bf Q}\cdot {\bf R}_i}\;
\langle n^\alpha_{i,\uparrow}-n^\alpha_{i,\downarrow} \rangle
\]
with $\alpha\in \{c,f\}$ and the quasiparticle gap
$\Delta_{QP}$ obtained from the bond Fermion formalism and QMC\cite{Assaad}.
Here we define $\Delta_{QP}$ as the energy between the highest
occupied and lowest unoccupied values of $E_{\nu,{\bf k}}$ so that the
numerical values should be twice the ones given in Ref. \cite{Assaad}.
When plotted versus $J/J_c$ the ordered moment agrees
reasonably well with the QMC results. 
In the QMC calculation the ordered moment
is obtained as the sqare root of the static structure factor
so that its sign is undetermined. The bond Fermion
calculation predicts the $f$ and $c$ ordered moments to have opposite 
sign - which appears plausible due to their
antiferromagnetic coupling - and we have assumed this to be true also for 
the QMC results. From their Dynamical Mean Field Theory calculations
Peters and Pruschke indeed found opposite direction of $m_c$ and
$m_f$ in Ref. \cite{PetersPruschke}.
The ordered moment of the $f$-electrons approaches the maximum value
of $0.5$ as $J/t\rightarrow 0$ whereas the $c$-electron ordered moment
approaches zero as $J/t\rightarrow 0$. This is to be expected
because the $c$-moment is reduced
by charge fluctuations and as $J\rightarrow 0$ the effective staggered field
due to the ordered $f$-spins vanishes.
The quasiparticle gap agrees reasonably well with the QMC result
and is roughly linear in $J$. In the bond Fermion result
there is a kink at $J=J_c$ which is not present in the
QMC data - on the other hand, using the dynamical cluster approximation
Martin {\em et al} found a very similar kink in the $\Delta_{QP}$
vs. $J$ curve\cite{MartinBerxAssaad}.
We proceed to the doped case. Figure \ref{fig5} shows the
energy as a function of $\Theta$ for different 
$\delta$ whereby $t_1/t=0.4$, and 
$J/t=1.4$ or $J/t=1.2$. As $\delta$ increases there appears -
in addition to the minimum for the Kondo insulator -
a `wiggle' in the $E$ vs. $\Theta$-curves which
develops into a second minimum.
For $J/t=1.4$ the lower of the two minima
shifts to smaller $\Theta$ with increasing $\delta$
and merges with the maximum at $\Theta=0$ into
a new minimum at this angle. This corresponds to a hole-doping
driven 2$^{nd}$ order
transition from the antiferromagnetic to the paramagnetic phase.
The second minimum - which always is higher in energy and thus
never realized - moves to slightly larger $\Theta$ and crosses
above the extremum at $\Theta=0$ well before the  2$^{nd}$ order
transition occurs.\\
This behavior changes for the value $J/t=1.2$.
The minimum for the Kondo insulator now shifts to larger values
of $\Theta$ as $\delta$ increases and crosses above the extremum
at $\Theta=0$ between $\delta=0.20$ and $\delta=0.24$ - which corresponds
to a 1$^{st}$ order transition. The second minimum still undergoes
the 2$^{nd}$ order transition but now is always higher in energy
and thus never realized. We therefore have two
antiferromagnetic phases with different behaviour upon increasing
doping. In the following we refer to the phase undergoing the
2$^{nd}$ order transition as AF I, the other one as AF II.\\
It is also apparent from Figure \ref{fig5} that the energy difference between
antiferromagnetic and paramagnetic phase at $\delta=0$ is only
$\approx 0.05\;t$ per site. The energy per site of the paramagnetic
phase itself at  $\delta=0$ and $J/t=1.4$
is $-1.82\;t$ so that the energy difference between the two phases
is only a few per cent of this value despite the fact that the
value of $J/t$ is already far below $J_c/t$. Obviously the relatively 
simple bond Fermion calculation does not reach such a level of accuracy 
so that the value of $J_c/t$ is off by a factor of $1.6$.\\
Proceeding as above we can map out the phase diagram in the $J-\delta$ 
plane which is shown in Figure \ref{fig6} for the two values
$t_1/t=0.4$ and $t_1/t=0.0$. The line separating the AF I and paramagnetic
phase corresponds to a 2$^{nd}$ order transition, the line
separating the AF II from either the AF I or the paramagnetic phase
represents a 1$^{st}$ order transition. The coexistence curves between
AF I and AF II do not reach the $\delta=0$ axis because
for very small doping there is only a single minimum in the 
energy-versus-$\Theta$ curves, see Figure \ref{fig5}.\\
Figure \ref{fig7} compares the phase diagram to that 
obtained by Watanabe and Ogata\cite{WatanabeOgata} using the
Variational Monte Carlo (VMC) method.
For $t_1=0$, VMC finds $J_c/t=1.7$, close to the exact result $J_c/t=1.45$
from QMC\cite{Assaad}. When plotting the phase diagram as function
of $J/J_c$ and $\delta$ as in Figure \ref{fig7} the bond Fermion
result agrees reasonably well with the VMC phase diagram. As will be discussed
below, the nature of the AF I and AF II phases also agrees with
VMC. A qualitatively similar phase diagram has also been obtained by
Lanata {\em et al.} using the Gutzwiller approximation in Ref. \cite{Lanata}
and by Asadzadeh {\em et al} also by using the VMC method\cite{Asadzadeh}.
To understand the nature of the two antiferromagnetic phases 
Figure \ref{fig8} shows their band structure and
momentum distribution function whereas Figure \ref{fig9}
shows the Fermi contours in the Brillouin zone.
For AF I the band structure may be thought of as having been obtained
from the paramagnetic one in Figure \ref{fig2} by antiferromagnetic folding
plus the formation of small `antiferromagnetic gaps' at the intersection
points of the original and folded bands.
Accordingly the Fermi surface is a hole pocket around $(\pi,\pi)$ plus its
antiferromagnetic umklapp around $(0,0)$. As can be seen
from the tiny discontinuities in $n_{\bf k}$ and the small slope
of the bands, the `heavy part' of the paramagnetic band persists
in this phase but undergoes antiferromagnetic folding.
This is quite different for AF II. There is still
a dispersionless band close to $E_F$ but the Fermi surface is now formed
by a more strongly dispersive band which has a strong $c$-character.
The Fermi surface now consists of a small pocket around
$(\frac{\pi}{2},\frac{\pi}{2})$ which may be thought of as having
being obtained by hybridizing the conduction electron
Fermi surface for electron density $n_c=0.9$ and its antiferromagnetic umklapp 
(for $t_1=0$ the Fermi surface in the AF-II phase really is identical to 
the noninteracting conduction electron Fermi surface plus its umklapp because 
in this case
there is no overlapp between original Fermi surface and umklapp).
The discontinuity in  $n_{\bf k}$ is much larger than for the AF I solution
at the side of the pocket facing $(0,0)$ whereas it is small at the part
of the pocket facing $(\pi,\pi)$ (as is also familiar from the spin-density
wave mean-field solution to the Hubbard model).
The plot of $n_{\bf k}$ also shows
why the AF II phase is favoured for small $J/t$:
the expectation value of the kinetic energy can be written as
$\langle H_t \rangle = 2\sum_{\bf k}\;\epsilon_{\bf k}\;n_{\bf k}$
and this is minimized by the noninteracting
function $n_{\bf k}^{(0)}=\Theta(E_F-\epsilon_{\bf k})$.
Figure \ref{fig8} shows that the phase AF II has an 
$n_{\bf k}$ which is closer to the noninteracting one and thus
has a lower kinetic energy than both the paramagnetic and the AF I phase.\\
A very similar classification of the AF I and AF II phase has been given by 
Watanabe and 
Ogata\cite{WatanabeOgata} based on their VMC calculations.
On the other hand, a quite different
phase diagram was obtained by by Martin and Assaad using
the dynamical cluster approximation\cite{MartinAssaad}.
This authors find only one antiferromagnetic phase with
a Fermi surface which does not include the $f$-electrons - which would
corresponding to the AF II solution above. Interestingly, the
band structure shown in Figure 5 of Ref. \cite{MartinAssaad}
also has some similarity with that of the AF II phase
in Figure \ref{fig8}. Martin and Assaad
find that the ordered moment vanishes continuously at the phase transition
between antiferromagnetic and paramagnetic phase
suggesting it to be 2$^{nd}$ order despite the discontinuous
change of the Fermi surface.\\
\begin{figure}
\includegraphics[width=0.7\columnwidth]{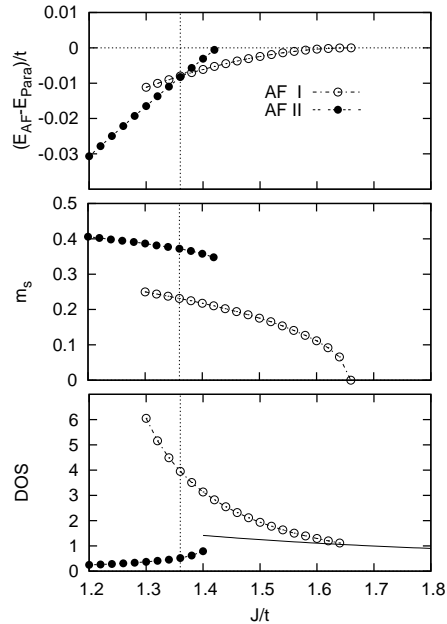}
\caption{\label{fig10} 
Ground state energy relative to the paramagnetic phase,
ordered $f$-like moment $m_s$ and density of states per lattice site, spin,
and energy $t$ at the Fermi level (DOS) as a function of $J/t$ for
$\delta=0.08$ and $t_1/t=0.4$. All data are shown for the AF-I and AF-II 
phases, the phase transition at $J/t=1.36$ is marked by the dashed vertical 
line. The full line in the panel for DOS is the density of states for the
paramagnetic solution.}
\end{figure}
Figure \ref{fig10} shows the development of some quantities
for fixed $\delta=0.08$ upon variation of $J/t$ from
$J/t=1.2$ and $J/t=1.8$. The value $t_1/t=0.4$ so that one should
compare to the upper panel of Figure \ref{fig6}.
At $J/t\approx 1.36$ the energies of the
AF-I and AF-II phases cross resulting in a
1$^{st}$ order phase transition.
The ordered $f$-moment drops by roughly $1/3$ at the transition,
whereas the density of states at the Fermi level  -
which is proportional to the effective mass - increases
significantly. At $J/t\approx 1.64$ the ordered moment
vanishes continuously at the  2$^{nd}$ order transition and the 
density of states at the Fermi level smoothly approaches that of the 
paramagnetic phase. Similar behaviour has been found
by Lanata {\em et al.} using a Gutzwiller wave function\cite{Lanata} and
more recently by Kubo using VMC for the periodic Anderson lattice\cite{Kubo}.
\section{Comparison to Experiment}
Let us next discuss the possible relevance of these results for 
experiments on $4f$-electron compounds. Here especially the
compounds CeCoIn$_5$, CeRhIn$_5$ and CeIrIn$_5$ come to mind.
These have a layered tetragonal crystal structure where CeIn planes
parallel to the $a$-$b$ plane and - say - CoIn$_4$ double layers are stacked 
alternatingly along the $c$-axis\cite{Shishido_1}. The Ce-ions in the 
CeIn planes
form a 2D square lattice as considered in the present calculation.\\
CeRhIn$_5$ is antiferromagnetic and the ordered moment withtin
a given CeIn-plane forms the simple two-sublattice N\'eel order
assumed in the present calculation. More precisely, there are several
antiferromagnetic phases all of which have simple N\'eel order
within the CeIn-planes, but differ in the component of the
magnetic ordering vector perpendicular to these.
The Fermi surface of CeRhIn$_5$ as measured by de Haas-van Alphen experiments 
appears to consist of several roughly cylindrical and quasi-2D sheets, 
the measured
cross sections at ambient pressure are very similar to those of 
LaRhIn$_5$\cite{Shishido_1} which implies that the Ce-4f electrons do not 
contribute to the Fermi surface volume. Pure CeRhIn$_5$ at ambient pressure
would thus correspond to the AF II phase discussed above
(the magnetic order is actually incommensurate in $c$-direction but
this is too subtle a detail for our highly simplified model anyway).
In contrast, the volume of the Fermi 
surfaces of CeCoIn$_5$ and CeIrIn$_5$ is consistent with itinerant Ce-4f 
electrons\cite{Shishido_1}.
The Fermi surface volume and 
magnetic structure CeRhIn$_5$ can be changed by either applying pressure
or by substituting Rh by Co or Ir. Thereby a  reduction of the lattice constant
by either applying pressure or by substituting Rh-ions by the isovalent
but smaller Co-ions
in the alloy CeRh$_{1-x}$Co$_x$In$_5$ has a different effect.\\
Alloying with Co in CeRh$_{1-x}$Co$_x$In$_5$ first - at $x\approx 0.4$ -
induces a phase transition between two antiferromagnetic phases,
from the incommensurate antiferromagnetic (ICAM) phase
phase with an incommensurate $c$-component of the
ordering vector to the commensurate antiferromagnetic (CAM) phase
with strict 3D N\'eel order\cite{Goh}.
The Fermi surface volume changes from CeRhIn$_5$-like
in the ICAM phase to  CeCoIn$_5$-like in the CAM phase so that the
transition from localized to itinerant Ce 4f electrons occurs at this
transition between the two antiferromagnetic phases and not
at the transition to the paramagnetic phase\cite{Goh}.
The ordered moment drops from $0.38\;\mu_B$/Ce in the ICAM phase
to $0.21\;\mu_B$/Ce in the CAM phase\cite{Yokohama}, the cylotron mass
increases from $4m_0$ to $10m_0$\cite{Goh}. It was moreover found 
that the N\'eel 
temperature is discontinuous across this transition which suggests it
to be 1$^{st}$ order\cite{Kawamura}. Increasing the doping to $x\approx 0.85$
induces a second phase transition to the paramagnetic phase
whereby neither the Fermi surface volume nor the cyclotron mass show
any discontinuity\cite{Goh}.\\
Identifying the ICAM phase with the
AF II phase and the CAM phase with the AF I phase and assuming that
alloying with Co simply amounts to increasing $J/t$ due to the contraction of
the lattice at constant $n_c$ all of this would
qualitatively match the behaviour in Figure \ref{fig10} quite well:
a 1$^{st}$ order transition from the AF II phase with large moment,
light effective mass and a Fermi surface which comprises only the
$c$-electrons to the AF I phase with small ordered moment,
large effective mass and a Fermi surface which comprises both $c$
and $f$-electrons, followd by a 2$^{nd}$ order transition
to the paramagnetic phase where neither the effective mass nor the
Fermi surface volume change. That the component
of the ordering vector perpendicular to the CeIn-planes
changes at the transition between AF II and AF I phase
would not be too surprising. If the hopping matrix elements
in $c$-direction change with the component of the
Bloch wave vector ${\bf k}$ within the CeIn-plane,
$t_z=t_z(k_x,k_y)$, a relatively strong change of the Fermi contour as in 
Figure \ref{fig9} certainly can strongly modify the coupling between the 
planes and thus affect the ordering vector.\\
On the other hand, applying pressure to CeRhIn$_5$ leads to a completely 
different behaviour. At $\approx 1GPa$ there is a first
transition - or rather a `crossover' - where only the component
of the magnetic ordering vector perpendicular to
the CeIn planes changes\cite{Raymond}. No change of either the
Fermi surface volume or the cyclotron mass  is observed at this
crossover\cite{Shishido_2}. In our simplified model a change of
the magnetic order perpendicular to the planes cannot be modelled
so we ignore this transition. One possible explanation for this
transition would be that the effective mass changes
with pressure - as seen in Figure \ref{fig10} if one assumes that
pressure changes $J/t$ - and if the interplane hopping $t_z$ is 
renormalized by the quasiparticle weight $Z \propto m_{eff}^{-1}$\cite{bilayer}
this may as well influence the coupling between planes.
Increasing the pressure further induces a second transition at $p_0=2.3 GPa$ to
a paramagnetic phase whose Fermi surface cross sections are
identical to those of CeCoIn$_5$ - corresponding to itinerant
Ce 4f electrons - whereas the cyclotron masses seem to
diverge at the transition\cite{Shishido_2}. The limiting values
of the cyclotron masses
approached for large $|p-p_0|$ on the two sides of the transition
differ considerably, whereby the values in the paramagnetic
phase are larger by a factor of $\approx 4$.
With the exception of the divergence of the cyclotron mass at the transition
- which cannot be reproduced by a simple theory as the present one -
this could correspond to a direct (1$^{st}$ order) transition AF-II 
$\rightarrow$ Para in the phase diagram in Figure \ref{fig6}.\\
Generally, reducing the volume of the lattice
by either applying pressure or isovalent doping with
ions with a smaller ionic raius may be expected to enhance the
overlap of atomic orbitals so that all hopping elements increase,
but the rate of increase varies with the
character of the two connected orbitals. On the other hand,
the intra-atomic Coulomb repulsion $U$ will remain essentially unchanged,
whereas the charge transfer energy $\epsilon_f$ may change due to a variation
of the Madelung potential although it is hard to predict
if $\epsilon_f$ increases or decreases.
From the Schrieffer-Wolff transformation the Kondo exchange constant
$J=2V^2((U+\epsilon_f)^{-1} + \epsilon_f^{-1})$ where $V$ is the $c$-$f$ hybridization.
Due to the higher power of $V$ in $J$, a plausible guess is
that $J/t$ increases upon contraction of the crystal.
On the other hand, this may not be the only effect of the contraction.
Since the hybridization integrals between different pairs of
orbitals change at a different rate, the width of different bands also
may change at a different rate. Accordingly it may happen that small
`uncorrelated' electron- or hole-pockets participating in the Fermi surface -
which are known to exist in CeRhIn$_5$ from de-Haas van-Alphen
experiments\cite{Shishido_1,Shishido_2} -
may shrink or expand if the band which
forms the pocket changes its width at a different rate
as compared to band which mixes with the
4f electrons. In this way, electrons
may be transferred from the pockets to the band which
interacts with the localized moments or vice versa so that
contraction of the lattice may in addition
change also the conduction electron density
$n_c$ in our simplified model. Having noticed this
we conclude that it may also make a difference if the
contraction is due to hydrostatic pressure or by doping the
material with ions that have a smaller ionic radius because
these perturbations may affect different bonds in the solid in a different
way. Therefore it might not be too surprising if applying hydrostatic 
pressure or alloying with Co would drive CeRhIn$_5$ through a phase 
diagram like that in Figure \ref{fig6} along different routes.\\
Another example for a  compound undergoing a pressure induced phase 
transition from an antiferromagnetic phase to a paramagnetic phase is
CeRh$_2$Si$_2$\cite{Araki} which also has layered structure comprising
planes where Ce atoms form a 2D square lattice.
At $p_0=1.1\;GPa$ antiferromagnetism
disappears and the Fermi surface changes from being consistent with
the density functional Fermi surface of LaRh$_2$Si$_2$ to
being consistent with that for CeRh$_2$Si$_2$ - which again corresponds to the
Ce-4f electrons changing from localized to itinerant\cite{Araki}.
Whereas three sheets of the Fermi surface could be resolved at ambient
pressure only one sheet is observed above $p_0$ and this has a cyclotron
mass which is larger by a factor of $2-4$ than the ones below $p_0$.
Again, there seems to be a transition from an antiferromagnetic
phase with localized $f$-electrons and relatively light masses
to a paramagnetic phase with itinerant $f$-electrons and heavy
masses which might correspond to the transition AF II $\rightarrow$
paramagnetic in the calculated phase diagram. 
Lastly, a transition between two different antiferromagnetic
phases has been observed in CeCu$_{6-x}$Au$_x$\cite{Hamann}.
Applying pressure to CeCu$_{5.5}$Au$_{0.5}$ results in an apparent
1$^{st}$ order transition between two phases which differ in their
ordering vector.
\section{On the Violation of the Constraint}
\begin{figure}
\includegraphics[width=0.7\columnwidth]{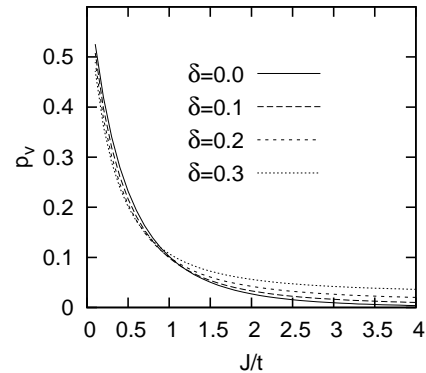}
\caption{\label{fig11} 
Probability $p_v$ for violation of the constraint
in the paramagnetic solution versus $J/t$ for different hole doping $\delta$.}
\end{figure}

Lastly we discuss the approximations that were made.
As already stated, the bond Fermions
in principle have to obey a hard-core constraint, that means no
two of them - which could differ either in their spin or
species - are allowed to occupy the same unit cell because such a state
cannot be translated meaningfully to a state of the true
Kondo lattice. Instead, in all the above calculations the bond Fermions
were treated as free Fermions which is an uncontrolled
approximation. However, there is a simple way to judge inhowmuch this 
assumption is justified namely to calculate for the ground state so obtained
the probability $p_v$ that the constraint is
violated at a given site. Since we are dealing with noninteracting
Fermions this is easily evaluated.
There are five allowed states of a given cell,
namely the empty cell or a state with
a single bond Fermion of either spin direction and either
species. The probability for the cell to be in one of these
five allowed states is
\begin{eqnarray*}
p_a&=&(1-n_a)^2\;(1-n_b)^2 + 2n_a\;(1-n_a)\;(1-n_b)^2 \nonumber \\
&&\;\;\;\;\;\;\;\;\;\;\;\;\;\;\;\; +
2n_b\;(1-n_a)^2\;(1-n_b)
\end{eqnarray*}
where $n_a=\langle \;a_{i,\sigma}^\dagger a_{i,\sigma}^{}\;\rangle$
and similar for $n_b$. 
The probability for the constraint to be violated then is
$p_v = 1 - p_a$.
Figure \ref{fig11} shows $p_v$ as a function of $J/t$ for different dopings.
For values of $J/t>1 $ one has $p_v\le 0.1$.\\
To put this in perspective, we consider other cases where a constraint
is relaxed. In linear spin wave theory for the spin-$\frac{1}{2}$ Heisenberg
antiferromagnet\cite{Anderson} the Bosonic magnons have to obey a hard-core
constraint because a given spin can be flipped only
once relative to the N\'eel order so that the presence of two magnons at the 
same site is unphysical. From the known reduction of the
ordered moment one can infer that even for the 2D case - where quantum 
fluctuations are strongest - the density of magnons is only $n_b=0.197$ 
per site. Accordingly, the probability that two
magnons occupy the same site - and thus violate the constraint -
is $p_v \approx n_b^2= 0.04$. In fact, linear spin wave
theory gives an excellent quantitative description of Heisenberg
antiferromagnets.\\
As a second example we consider mean-field theories for the Kondo
lattice\cite{LacroixCyrot,Lacroix,ZhangYu,Senthil}.
There the Heisenberg exchange is Hartree-Fock decoupled
resulting in a mean-field Hamiltonian which is a quadratic form 
and describes the
mixing between the original conduction band and a dispersionless
band of $f$-electrons. The number of $f$-electrons then is adjusted
to one per unit cell by tuning the energy of the effective $f$-level.
Since the $f$-electrons are uncorrelated and their density is $0.5$ per
unit cell and spin, the probability that there are either
$2$ or $0$ $f$-electrons 
in a cell so that the constraint is violated is $p_v=0.5$\\
For the bond Fermions the probability for a violation of the constraint 
thus is not as small as in linear spin wave theory and one may
expect stronger deviations e.g. for the ground state energy which
probably is the reason for the incorrect value of $J_c/t$.\\
One might consider an approximate treatment of the hardcore constraint
by Gutzwiller projection or by a mean-field theory similar to
the one proposed by Sachdev and Bhatt\cite{SachdevBhatt}.
In fact, using the mean-field procedure for the Kondo insulator with
$t_1=0$ Jurecka and Brenig obtained the value
$J_c/t=1.5$, very close to the exact value from QMC\cite{JureckaBrenig}.
However, we have tried both methods and abandoned them because 
both of them lead to a substantial narrowing of the conduction band.
On the other hand, all available numerical results for the single
particle spectral function of the Kondo lattice - or the related
periodic Anderson model - agree in that the conduction electron bandwith
retains its original width of $2zt$ (with $z$ the number of
nearest neighbors) see for example Fig. 1 in Ref. \cite{Tsutsui},
Fig. 3 in Ref. \cite{Oana}, or Fig. 4 in Ref \cite{MartinAssaad}.
Any `correlation narrowing' of the conduction
band thus obviously is unphysical. Moreover, as was discussed above,
the bond Fermion theory without any constraint
does reproduce the correct limiting value of the ground state
energy for $J/t\rightarrow 0$. This favourable property is lost if
any renormalization of the hopping integrals is introduced.
The best procedure therefore probably is to accept the inaccuracy and 
simply relax the constraint without further correction.
\section{Conclusion}
In summary we have presented a theory for ground state and single particle 
spectrum the Kondo lattice based on bond Fermions. Thereby the constraint of 
having precisley one $f$-electron per unit cell is fulfilled
exactly, instead we are relaxing the hard core constraint on the
bond Fermions which however turns out to be reasonably justified as
the probability for violation of the constraint is low for not
too small values of $J/t$. While being of comparable simplicity as mean-field 
theory the bond Fermion theory gives results which
differ significantly from mean-field theory. Rather, the theory reproduces
qualitatively a number of results obtained previously only by numerical 
methods. In particular we find two different antiferromagnetic phases, the 
first one with a small ordered moment and antiferromagnetically folded `heavy' 
bands where the localized electrons do contribute to the Fermi surface volume,
the second one with larger ordered moment and a Fermi surface which corresponds
to the backfolded Fermi surface of the conduction electrons alone and 
relatively light bands at the Fermi surface. Qualitatively the resulting phase 
diagram is quite consistent with experiments on CeRhIn$_5$.
While in the present manuscript we have studied only simple two-sublattice 
N\'eel order the generalization to more complicated magnetic 
structures is self-evident.


\begin{thebibliography}{}
\bibitem{Review}
H. v. L\"ohneysen, A. Rosch, M. Vojta, and P. W\"olfle, 
Rev. Mod. Phys. {\bf 79}, 1015 (2007).
\bibitem{SchriefferWolff}
J. R. Schrieffer and P. A. Wolff, Phys. Rev. {\bf 149}, 491 (1966).
\bibitem{Doniach}
S. Doniach, Physica B {\bf 91}, 231 (1977).
\bibitem{YoshimoriSakurai}
A. Yoshimori and A. Sakurai,
Progr. Theor. Phys. Supp. {\bf 46}, 162 (1970).
\bibitem{LacroixCyrot}
C. Lacroix and M. Cyrot, Phys. Rev. B {\bf 20}, 1969 (1979).
\bibitem{Lacroix}
C. Lacroix,
Journal of Magnetism and Magnetic Materials {\bf 100},  90 (1991).
\bibitem{ZhangYu}
G.-M. Zhang and L. Yu, 
Phys. Rev. B 62,  {\bf 76} (2000).
\bibitem{Senthil}
T. Senthil, M. Vojta, and S. Sachdev,
Phys. Rev. B {\bf 69}, 035111 (2004).
\bibitem{ZhangSuLu}
G.-M. Zhang, Y.-H. Su, and L. Yu,
Phys. Rev. B {\bf 83}, 033102 (2011).
\bibitem{Nilsson}
J. Nilsson, Phys. Rev. B {\bf 83}, 235103 (2011).
\bibitem{Asadzadeh}
M. Z. Asadzadeh, F. Becca, and M. Fabrizio,
Phys. Rev. B {\bf 87}, 205144 (2013).
\bibitem{renormalization}
S. J. Yamamoto and Q. Si, 
Phys. Rev. Lett. {\bf 99}, 016401 (2007).
\bibitem{EDMFT}
Q. Si, J. H. Pixley, E. Nica, S. J. Yamamoto,
P. Goswami, R. Yu, and S. Kirchner,
J. Phys. Soc. Jpn. {\bf 83}, 061005 (2014).
\bibitem{Global}
M. Vojta, Phys. Rev. B {\bf 78},  125109 (2008).
\bibitem{woelfle}
E. Abrahams, J. Schmalian, and P. W\"olfle,
 Phys. Rev. B {\bf 90}, 045105 (2014).
\bibitem{twofluid}
G. Lonzarich, D. Pines, and Y.F. Yang,
preprint arXiv:1601.06050.
\bibitem{SachdevBhatt}
S. Sachdev and R.N. Bhatt, Phys. Rev. B {\bf 41}, 9323 (1990).
\bibitem{Gopalan}
S. Gopalan, T. M. Rice, and M. Sigrist,
 Phys. Rev. B {\bf 49}, 8901 (1994).
\bibitem{Oana}
R. Eder, O. Stoica, and G. A. Sawatzky,
Phys. Rev. B. {\bf 55}, R6109 (1997);
R. Eder, O. Rogojanu, and G. A. Sawatzky,
Phys. Rev. B. {\bf 58}, 7599 (1998).
\bibitem{JureckaBrenig}
C. Jurecka and W. Brenig, Phys. Rev. B {\bf 64}, 092406 (2001).
\bibitem{YuWhite}
 C. C. Yu and S. R. White, Phys.\ Rev.\ Lett. {\bf 71}, 3866 (1993).
The DMRG data for the ground state energy are not published in this reference 
but seem to be given only in Ref. \cite{series}.
\bibitem{series}
Z.-P. Shi, R. R. P. Singh, M. P. Gelfand, and Z. Wang
Phys. Rev. B {\bf 51}, 15630(R) (1995).
\bibitem{Assaad}
F. F. Assaad, Phys. Rev. Lett. {\bf 83}, 796 (1999).
\bibitem{PetersPruschke}
R. Peters and T. Pruschke, Phys. Rev. B {\bf 76}, 245101 (2007).
\bibitem{MartinBerxAssaad} 
L. C. Martin, M. Bercx, and F. F. Assaad,
Phys. Rev. B {\bf 82}, 245105 (2010).
\bibitem{WatanabeOgata}
H. Watanabe and M. Ogata, Phys. Rev. Lett. {\bf 99}, 136401 (2007).
\bibitem{Lanata}
N. Lanata, P. Barone, and M. Fabrizio,
Phys. Rev. B {\bf 78}, 155127 (2008).
\bibitem{MartinAssaad}
L. C. Martin and F. F. Assaad,
Phys. Rev. Lett. {\bf 101}, 066404 (2008).
\bibitem{Kubo}
K. Kubo, J. Phys. Soc. Jpn. {\bf 84},  094702 (2015).
\bibitem{Shishido_1}
H. Shishido, R. Settai, D. Aoki, S. Ikeda, H. Nakawaki, N. Nakamura, 
T. Iizuka, Y. Inada, K. Sugiyama, T. Takeuchi, and K. Kin,
J. Phys. Soc. Jpn. {\bf 71}, 162 (2002).
\bibitem{Goh}
S. K. Goh, J. Paglione, M. Sutherland, E. C. T. O’Farrell, C. Bergemann, 
T. A. Sayles, and M. B. Maple,
Phys. Rev. Lett. {\bf 101}, 056402 (2008)
\bibitem{Yokohama}
 M. Yokoyama, H. Amitsuka, K. Matsuda, A. Gawase, N. Oyama, I. Kawasaki, 
K. Tenya, and H. Yoshizawa,
J. Phys. Soc. Jpn. {\bf 75},  103703 (2006).
\bibitem{Kawamura}
S. Ohira-Kawamura, H. Shishido, A. Yoshida, R. Okazaki, H. Kawano-Furukawa, 
T. Shibauchi, H. Harima, and Y. Matsuda,
Phys. Rev. B {\bf 76}, 132507 (2007).
\bibitem{Raymond}
S. Raymond, G. Knebel, D. Aoki, and J. Flouquet
Phys. Rev. B {\bf 77}, 172502 (2008).
\bibitem{Shishido_2}
H. Shishido, R. Settai, H. Harima, and Y. Onuki,
J. Phys. Soc. Jpn. {\bf 74},  1103 (2005).
\bibitem{bilayer}
R. Eder, Y. Ohta, and S. Maekawa, Phys. Rev. B {\bf 51}, 3265 (1995).
\bibitem{Araki}
S. Araki,  R. Settai, T. C. Kobayashi, H. Harima, and Y. Onuki,
Phys. Rev. B {\bf 64}, 224417 (2001).
\bibitem{Hamann}
A. Hamann, O. Stockert,  V. Fritsch,  K. Grube,  A. Schneidewind,  
and H. v. L\"ohneysen,
Phys. Rev. Lett. {\bf 110}, 096404 (2013).
\bibitem{Anderson}
P. W. Anderson, Phys. Rev. {\bf 86} 694 (1952).
\bibitem{Tsutsui}
K. Tsutsui, Y. Ohta, R. Eder, S. Maekawa, E. Dagotto, 
and J. Riera, Phys. Rev. Lett.  {\bf 76}, 279 (1996).
\end{thebibliography}
\end{document}